\documentclass[aip,graphicx]{revtex4-1}
\usepackage{amsmath}  
\usepackage{graphicx}
\usepackage{color}
\newcommand{\bi}[1]{\ensuremath{\boldsymbol{#1}}}
\draft % marks overfull lines with a black rule on the right

\begin{document}

% on the title page in prepri mode.
%\preprint{}

\title{{Entrainment of noise-induced and limit cycle oscillators under weak noise}}




\author{Namiko Mitarai$^1$, Uri Alon$^2$, and Mogens H. Jensen$^1$}
%\thanks{}
\affiliation{$^1$Niels Bohr Institute, University of Copenhagen,
Belgdamsvej 17, Copenhagen, 2100-DK, Denmark.
\\
$^2$
Weizmann Institute of Science,
234 Herzl St.,
Rehovot 76100, Israel }


\date{\today}

\begin{abstract}
Theoretical models that describe
oscillations in biological systems are often
either a {\it limit cycle oscillator},
where the deterministic nonlinear dynamics
gives sustained periodic oscillations,
or a {\it noise-induced oscillator}, where
a fixed point is linearly stable
with complex eigenvalues and addition of
noise gives oscillations around the fixed point
with fluctuating amplitude. We investigate how each
class of model behaves under the external periodic
forcing, taking the well-studied van der Pol equation
as an example.
We find that, when the forcing is additive, the noise-induced oscillator
can show only one-to-one entrainment to the external frequency,
in contrast to the limit cycle oscillator which is known to entrain
to any ratio.  
When the external forcing is multiplicative,
on the other hand, the noise-induced oscillator can show entrainment
to a few ratios other than one-to-one, while the limit cycle oscillator shows
entrain to any ratio.  
The noise blurs the entrainment in general,
{ but clear entrainment regions for limit cycles can be identified 
as long as the noise is not too strong.}
\end{abstract}

\pacs{05.45.Xt,05.40.Ca,87.10.Ed}% insert suggested PACS numbers in braces on next line

\maketitle %\maketitle must follow title, authors, abstract and \pacs

{\bf 
Biological systems present us with a wide range of oscillators, which include cell
cycles, circadian rhythms, calcium oscillations, pace maker cells, and protein
responses, but it is often a challenging task to identify the minimal models
behind these oscillations.  The proposed models are typically categorized into
two classes:
(i) {\it Limit cycle oscillator}, where
 fixed points are linearly unstable and the oscillations are described by
stable limit cycles sustained by nonlinearities of the system which are
deterministic. Noise can be added on the top of the deterministic
oscillations.
(ii) {\it Noise-induced oscillator}, where
the fixed point is linearly stable for the system without noise and the system
relaxes to the fixed point with damped oscillations when temporally perturbed.
Addition of noise to this type of system is known to show sustained oscillations with
fluctuating amplitudes. We propose a way to distinguish the two, by using the
phenomenon of entrainment to a periodic perturbation.  Taking the van der Pol
equation with noise as an example, we show that entrainments to all the
rational ratios are seen only in the  limit cycle oscillator. In the case of
the noise-induced oscillator with additive external forcing, the oscillator can
entrain only at one-to-one ratio, meaning that the entrainment to other than
the one-to-one ratio is the sign of the dominance of the limit cycle mechanism.
When the external forcing is multiplicative, we find that the noise-induced
oscillator with weak nonlinearity can show some entrainment ratios other than
one-to-one, but not all the ratios.
}

\section{Introduction}
Biological systems present us with
a bewildering fauna of oscillators: cell cycles \cite{Ferrell},
circadian rhythms \cite{Lefranc,Asher,Pfeuty, Gerard}, calcium
oscillations \cite{Goldbeter}, pace maker cells \cite{pacemaker}, protein
responses \cite{Hoffmann02,Nelson04,Krishna, Tiana, Geva-Zatorsky06,
Geva-Zatorsky,Lang},
and so on.
Sometimes, however, it is hard to
see what are the minimal models behind these oscillations.
Typically, the models are categorized into two classes:
(i) {\it Limit cycle oscillator}:
The fixed point is linearly unstable and
the oscillations are
described by stable limit cycles sustained by
nonlinearity of the system  
in the deterministic case \cite{Tiana,Geva-Zatorsky06}.
Noise (e.g. molecular noise due to limited number of copy numbers)
can be added on the top of the deterministic oscillations.
(ii) {\it Noise-induced oscillator}:
The fixed point is linearly stable for the system without noise
and the system relaxes to the  
fixed point with damped oscillations when temporally perturbed.
Addition of noise to such a system is known to
show sustained oscillations with fluctuating amplitude
\cite{Geva-Zatorsky,Lang}.
For some systems, both limit cycle oscillators (i)
and noise-induced oscillators (ii)
are proposed as a mechanism for the oscillation
\cite{Tiana,Geva-Zatorsky06,Geva-Zatorsky}. Here, we propose a way to distinguish the two,
by using the
phenomenon of {\it entrainment} to a periodic perturbation.

It is well known that, when an periodic perturbation is
added to a deterministic limit cycle,
the system's oscillation frequency $\omega$ will be entrained to
the external frequency $\Omega$
with various rational numbers of frequencies $\omega/\Omega=P/Q$
for {\it all} positive integers $P$ and $Q$
in a {\it finite window of the external frequency} $\Omega$,
where the width of the window depends on the amplitude of the
external forcing \cite{JBB1,JBB2}. Entrainment, also called
mode-locking, has been observed in
variety of physical systems
during the last decades, from onset of
turbulence \cite{Stavans}, Josephson junctions \cite{He,He2},
one-dimensional conductors \cite{Gruner}, semiconductors
\cite{Gwinn,Lindsey} and crystals \cite{Martin}. It has been
predicted, and verified experimentally, that the mode-locking structure
possesses certain universal properties \cite{JBB1,JBB2}.
In biological systems,  entrainment has  been
investigated theoretically for circadian rhythms \cite{Pfeuty, Gerard}
as well as in model systems for protein responses \cite{JK}.
Experimental observation of entrainment in biological systems is often rather difficult
due to noisy signals, but it
has been observed for circadian rhythms \cite{Lefranc,Asher}
{and synthetic genetic oscillators \cite{Hasty}}.

In this paper, we study the difference in the entrainment
behavior for the limit cycle oscillators and noise-induced oscillators.
{ Our main question is the following: Can we distinguish the two cases 
by means of the entrainment behavior?
We employ the famous van der Pol equation with noise as an example,
because there we can easily study both cases by changing parameters.}
 { We show that entrainments to all the
rational ratios are seen only in the  limit cycle oscillator. In the case of
the noise-induced oscillator with additive external forcing, the oscillator can
entrain only at one-to-one ratio, meaning that the entrainment to other than
the one-to-one ratio is the sign of the dominance of the limit cycle mechanism.
When the external forcing is multiplicative, we find that the noise-induced
oscillator with weak nonlinearity can show some entrainment ratios other than
one-to-one, but not all the ratios.
}
{ To confirm the generality of the
entrainment behavior for the limit cycle system under weak noise,
we also study a biological example,
the TNF-driven oscillating NF-$\kappa$B system, and
confirm that $P/Q$ entrainments can be seen.
}

\section{Model}
%\subsection{Two ways to obtain the oscillation}
%We consider the $d$-dimensional
%deterministic dynamical equations of the form
%\begin{equation}
%\dot{\bi x}(t)={\bi F}
%({\bi x}(t); \bi a).
%\end{equation}
%$\bi F({\bi x}(t); a)$
%is a nonlinear function of
%$\bi x$, and $a$ is a parameter set of the system.
%The fixed point ${\bi x_0^{a}}$
%for a given $a$ is determined by
%as
%\begin{equation}
%{\bi F}({\bi x_0^{a}}; a)=0.
%\end{equation}
%The linear stability of the fixed point
%is detemined by the Jacobian
%\begin{equation}
%J=\nabla {\bi F}({\bi x}; a)|_{\bi x = \bi x_0^{a}}.
%\end{equation}
%We assume that the system undergoes the Hopf bifurcation
%with changing the parameter $a$.
%
%We then add noise to the system to
%have the Langevin equation of the form
%\begin{equation}
%\dot{\bi x}(t)={\bi F}
%({\bi x}(t);a)+\sigma{\bi R(t)}
%\end{equation}
%where $\bi R(t)$ is $d-$dimensional
%Gaussian white noise satisfing
%\begin{equation}
%\langle R_i(t)\rangle=0,\quad
%%\end{equation}
%%\begin{equation}
%\langle R_i(t)R_j(t')\rangle=\delta_{i,j}\delta(t-t').
%\end{equation}

\subsection{van der Pol equation}
Consider the following two-dimensional equation
with noise:
\begin{equation}
\dot{\bi{x}}=\bi F(\bi x)+\sigma \bi \Gamma,
\end{equation}
with
\begin{equation}
\bi x=\left(
\begin{array}{c}
x_1(t)\\
x_2(t)
\end{array}
\right),
\quad
\bi \Gamma=\left(
\begin{array}{c}
\Gamma_1(t)\\
\Gamma_2(t)
\end{array}
\right),
\end{equation}
\begin{equation}
\bi F(\bi x)=
\left(
\begin{array}{c}
x_2(t)
\\
-(Bx_1(t)^2-d)x_2(t)-x_1(t)
\end{array}
\right).
\end{equation}
Here, $d$, $\sigma$, and $B$ are parameters,
and $\Gamma_i(t)$ are  uncorrelated, statistically independent Gaussian
white noise, satisfying
\begin{eqnarray}
\langle \Gamma_j(t)\rangle=0, \quad
\langle \Gamma_j(t)\Gamma_k(t')\rangle=\delta_{j,k}\delta(t-t').
\end{eqnarray}

First let us consider the deterministic case, $\sigma=0$.
The model has a fixed point at $(x_1,x_2)=(0,0)$, and
the eigenvalues around this fixed point are
\begin{equation}
\lambda_{\pm}=\frac{1}{2}\left(d\pm \sqrt{d^2-4}\right),
\end{equation}
indicating that the system experiences a Hopf bifurcation at $d=0$.
When $d<0$, the fixed point
relaxes to the fixed point with damped oscillation
with the angular frequency $\omega_{\ell}(d)=\sqrt{|d^2-4|}/2$,
while when $d>0$ and $B>0$ the model shows a stable limit cycle
(van der Pol oscillator).

In the stochastic case with $\sigma>0$, however, the system
shows a sustained oscillation even in the linearly stable case,
$d<0$, because the noise keeps activating
the oscillation with frequency $\omega_{\ell}$.
This is the case of the linear p53 model
introduced in Ref. \cite{Geva-Zatorsky}.
When $d>0$,  $\sigma>0$ adds fluctuations on top of the stable
oscillation around the limit cycle.

\subsection{Setup}
We investigate the entrainment behavior of the model,
focusing  on the following three classes of parameter sets.
\begin{enumerate}
\item The {\it limit cycle oscillator},  
with $d>0$ and $B>0$.
\item For the {\it noise-induced oscillator},
we consider the two subcategories.
\begin{enumerate}
\item The {\it linear system with a stable fixed point},
with $d<0$ and $B=0$, i.e., the equations  
are linear in $\bi x$ and the fixed point is stable.

\item The {\it nonlinear system with a stable fixed point},
with $d<0$ and $B>0$, i.e.,
the fixed point is linearly stable but
the equations possess a nonlinear term.
\end{enumerate}
\end{enumerate}
When noise-induced oscillators are studied,
normally only linear terms are considered.
However, in reality, there are often nonlinear
terms, which can play a role when distance
from the stable fixed point $|\bi x|$
is sufficiently large. This is the reason why we consider
both linear and nonlinear noise-induced oscillators.

When needed, numerical integration of stochastic differential equations are performed by using Euler method.

\begin{figure*}[H]
\includegraphics[width=\textwidth]{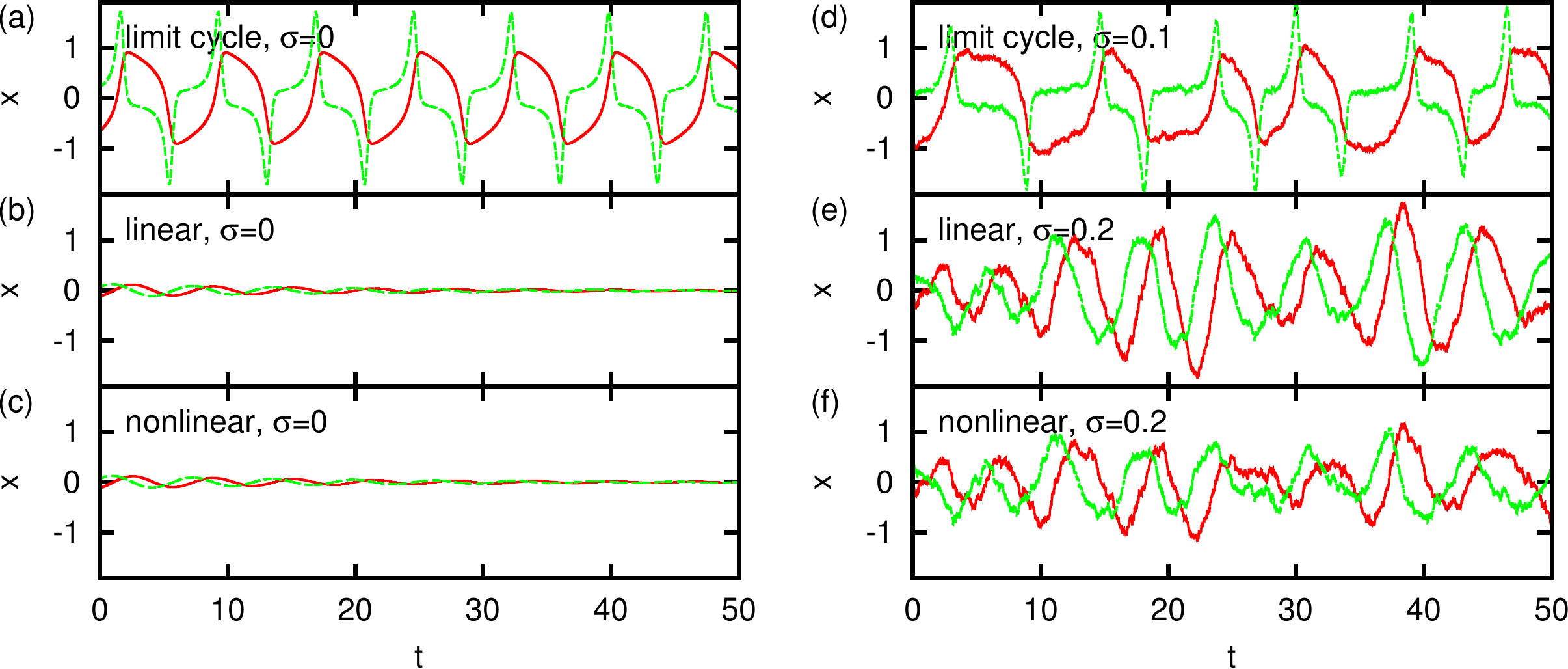}
\caption{(color online)
The time evolution of $x_1$ (solid line) and $x_2$ (dashed line)
when there is no external forcing.
(a) limit cycle oscillator with $d=2$ and $B=10$ without noise
($\sigma=0$).
(b) linear system with $d=-0.1$ and $B=0$ without noise
($\sigma=0$).
(c) nonlinear system with $d=-0.1$ and $B=1$ without noise
($\sigma=0$).
For (b) and (c), the initial condition is perturbed from the fixed point
to demonstrate the dumped oscillation.
(d) limit cycle oscillator with $d=2$ and $B=10$ with noise
($\sigma=0.1$).
(b) linear system with $d=-0.1$ and $B=0$ with noise
($\sigma=0.2$).
(c) nonlinear system with $d=-0.1$ and $B=1$ with noise
($\sigma=0.2$).
}
\label{noforce}
\end{figure*}
Figure \ref{noforce} shows the typical behavior of the
model in each categories. The parameters are chosen so that
the period and amplitude are in similar range.
Without noise, the limit cycle is the only case with stable oscillation
(Fig. \ref{noforce}a),
while linear and nonlinear systems with  a stable fixed point
exhibit damped oscillations relaxing to the fixed point
(Fig. \ref{noforce}bc).
When noise is added, the oscillation is perturbed
for limit cycle oscillator (Fig. \ref{noforce}d);
here the noise level is chosen so that the base oscillation
is still recognizable.
For linear and nonlinear noise-induced oscillators (Fig. \ref{noforce}ef),
we observe oscillations with the expected angular frequency
($\omega_{\ell}(-0.1)\approx 1$).
In order to demonstrate the difference
between the two, we apply the exact same sequence of noises
in both cases. We observe a bigger difference when
linear noise-induced oscillator have large ($|\bi x|\approx 1$)
amplitude, because the nonlinear term becomes more important.
Naturally this effect depends on the value of $B$ (data not shown).

We study these oscillators under the following
two kinds of external periodic perturbation.
\paragraph{Additive forcing.}
The first case is an {\sl additive} forcing,
in the form of
\begin{equation}
\dot{\bi{x}}=\bi F(\bi x)+
\sigma \bi \Gamma+\bi A(t),
\end{equation}
with
\begin{equation}
\bi A(t)= \left(
\begin{array}{c}
0\\
A
\end{array}
\right)
\cos\Omega t,
\end{equation}
\paragraph{Multiplicative forcing.}
The second case is an {\sl multiplicative} forcing (also called
parametric forcing),
in the form of
\begin{equation}
\dot{\bi{x}}=\bi F(\bi x)+
\sigma \bi \Gamma+\mathbf M(t)\bi x,
\end{equation}
where
\begin{equation}
\mathbf M(t)=\left(
\begin{array}{c c}
0 & 0\\
M& 0
\end{array}
\right) \cos\Omega t.
\label{whatisM}
\end{equation}

In the next section, we first present the behavior of
the model under the additive forcing, and
then show the parallel  results for the multiplicative forcing.

\section{Results}
\subsection{Additive forcing}
\subsubsection{Linear case}
%In the deterministic case with a linearly stable fixed point,
%the long-time behavior of the linearized equations
%around the fixed point under the external forcing
%can be analytically evaluated.  for any dimension. %as described in the following.

In the case of the additive periodic forcing to a linear deterministic system,
we have in general
\begin{equation}
\dot{\bi x}(t)=\mathbf L{\bi x}(t)+\bi A(t),
\label{linearsolve}
\end{equation}
where $\mathbf L$ is a coefficient matrix of the linearized equation,
and $\bi A(t)$ is periodic function in time with a period $T$,
satisfying $\bi A(t+T)=\bi A(t)$.

{
By expressing $\bi x(t)=\sum_{j=1}^d C_j(t)\bi u_j$,
with using eigenvectors $\bi u_j$ of the
matrix $L$ given by $\mathbf L \bi u_j=\lambda_j \bi u_j$,
we can show that in the long-time limit we have
\begin{eqnarray}
\lim_{t\to \infty} C_j(t)
=\sum_{n=-\infty}^{\infty}
\frac{F_n }{i n\frac{2\pi}{T}-\lambda_j}
e^{in\frac{2\pi}{T} t}.
\end{eqnarray}
Note that $\Re(\lambda_j)<0$ because
the fixed point $x=0$ is stable.
$F_n$ is defined by the Fourier expansion of $\bi A(t)$ as
\[
\bi v_j^t\cdot \bi A(t)=\sum_{n=-\infty}^{\infty}F_n
e^{i n \frac{2\pi}{T} t},
\]
where $\bi v_j^t$ is the left eigenvector.
%in the sense that
%$\bi v_j^t\mathbf L=\lambda_j\bi v_j^t$
%with normalisation
%$\bi v_j^t\cdot \bi u_k^t=\delta_{j,k}$.
Therefore the solution will always be a periodic function of $t$
with the period $T$ in the long time limit,
and contains only the frequencies that the external forcing has.
In other words, the system will be
always in a 1/1 entrained state  
if the perturbation is pure sine or cosine wave.
}

{
%\subsubsection{Linear case with noise}
When Gaussian white noise is added to eq. (\ref{linearsolve}), we have
\begin{equation}
\dot{\bi x}(t)=\mathbf L{\bi x}(t)+\bi A(t) + \sigma\bi\Gamma(t).
\label{linearsolve}
\end{equation}
In this case, we can evaluate the auto-correlation of $C_j(t)$ 
%similar calculation as no-noise case gives
%\begin{equation}
%C_j(t)=C_0 e^{\lambda_j t}
%+\sum_{n=-\infty}^{\infty}
%\frac{F_n }{i n\frac{2\pi}{T}-\lambda_j}\left[
%e^{in\frac{2\pi}{T} t}-e^{\lambda_jt}\right]
%+\int_0^t\sigma R_j(s)e^{-\lambda_j (s-t)}ds,
%\end{equation}
%with
%\[
%R_j(t)\equiv \bi v_j^t\cdot \bi \Gamma(t).
%\]
%The  auto-correlation 
for large enough $t_0$ ($t_0>> 1/|\Re{\lambda_j}|$) as
\begin{equation}
%\lim_{t_0\to \infty}
\langle C_j(t_0)C_j(t_0+\tau)\rangle
\approx \sum_{n=-\infty}^{\infty}\sum_{n'=-\infty}^{\infty}
\frac{F_nF_{n'} }
{(i n\frac{2\pi}{T}-\lambda_j)(i n'\frac{2\pi}{T}-\lambda_j)}
e^{i(nt_0+n'(t_0+\tau))\frac{2\pi}{T}}
-\frac{\sigma^2}{2\lambda_j}e^{-\lambda_j \tau}.
\label{AC}
\end{equation}
Namely,  the response contains oscillations with 
the frequencies from  forcing $2\pi/T$
and from the complex part of the eigenvalue $\Im \lambda_j$, 
and the amplitude of the latter is proportional to $\sigma$.
}

\subsubsection{Numerical results}
We now investigate numerically the entrainment behaviors
for all three categories.
First we demonstrate the behavior without noise,
and then show how the noise modify this behavior.
\paragraph{Without noise.}
\begin{figure*}[H]
\includegraphics[width=\textwidth]{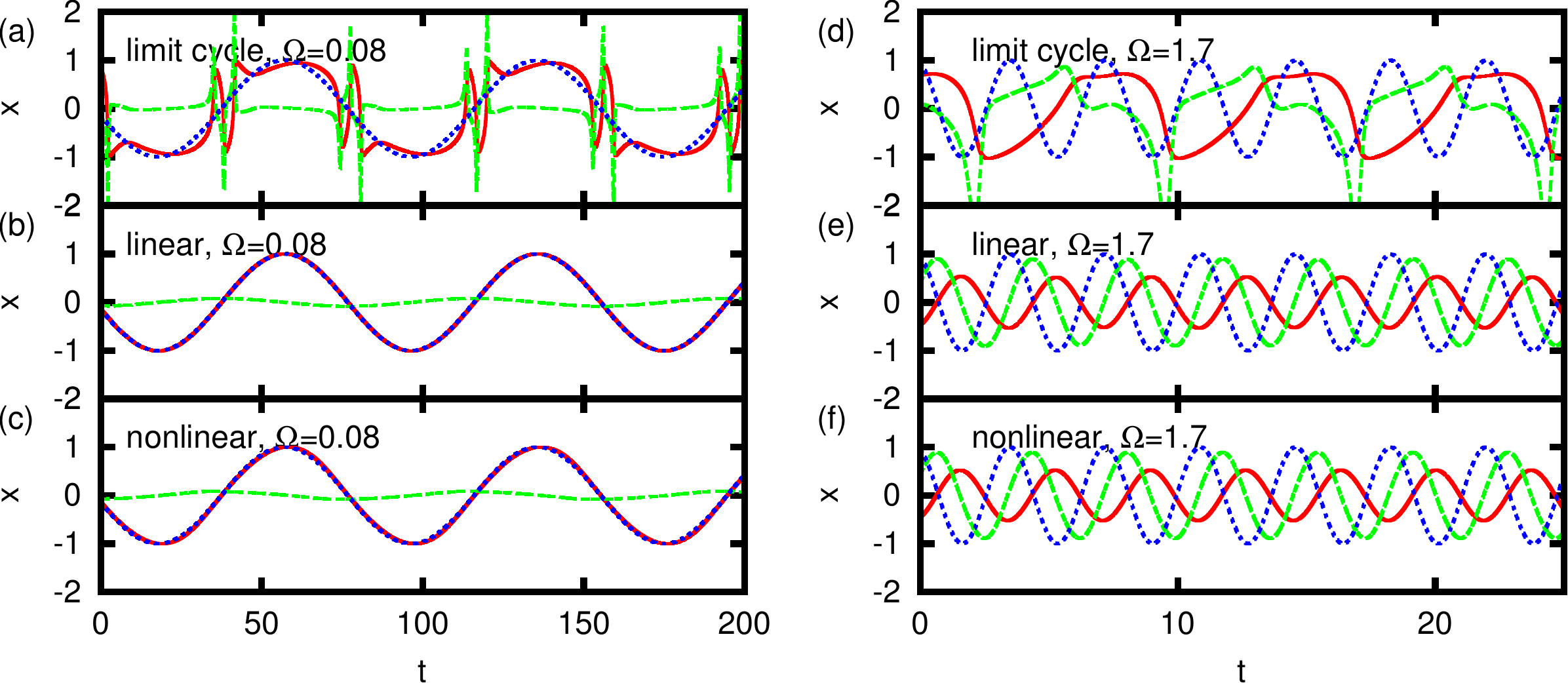}
\caption{(color online)
The time evolution of $x_1$ (solid line) and $x_2$ (dashed line)
when there is additive external forcing (dotted line, $A=1$).
The external forcing has angular frequency $\Omega=0.08$
for (a-c), and  $\Omega=1.7$  for (d-f).
(a) and (d) limit cycle oscillator with $d=2$ and $B=10$ without noise
($\sigma=0$).
(b) and (e) linear system with a stable fixed point with $d=-0.1$ and $B=0$ without noise
($\sigma=0$).
(c) and (f) nonlinear system with a stable fixed point with $d=-0.1$ and $B=1$ without noise
($\sigma=0$).
For the case with a limit cycle oscillator
(a, d), the system's angular frequency
can entrain to the external angular $\Omega$ with various ratios, while
in the linear and nonlinear systems with a stable fixed point case (b,c,e,f)
the system can only entrain to one to one ratio.
}
\label{AforceExample}
\end{figure*}

Figure \ref{AforceExample} illustrates typical entrainment behaviors
for additive forcing when noise is absent.
With a limit cycle oscillator
(Fig. \ref{AforceExample} a, d), the system's angular frequency
can entrain to the external angular $\Omega$ with various ratios, while
in the linear system, one-to-one entrainment occurs
(Fig. \ref{AforceExample} b, e).
The nonlinear system shows very similar behavior to
the linear system, where we see only one-to-one
entrainment (Fig. \ref{AforceExample} c, f).

In order to define the system's angular frequency in a
simple way, we adopt the polar coordinate $(r,\theta)$ using
\begin{eqnarray}
x_1(t)&=&r(t)\cos\theta(t),\\
x_2(t)&=&r(t)\sin\theta(t),
\end{eqnarray}
as proposed in Ref. \cite{Parlitz}. We define $\theta(t)$ so that
$(\theta(t)-\theta(0))/2\pi$ gives the winding number, i.e.,
how many times the orbit went around the fixed point by time $t$.
The system's angular frequency is numerically calculated from
\begin{equation}
\omega=\frac{1}{T}\left[\theta(T)-\theta(0)\right]
\end{equation}
for long enough $T$ (typically 1000 times external forcing period).
With this definition, Fig. \ref{AforceExample}(a)
shows the entrainment of the ratio
$\omega/\Omega=2/1$, while Fig. \ref{AforceExample}(d)
gives $\omega/\Omega=1/2$.

\paragraph{With noise.}
\begin{figure*}[H]
\includegraphics[width=\textwidth]{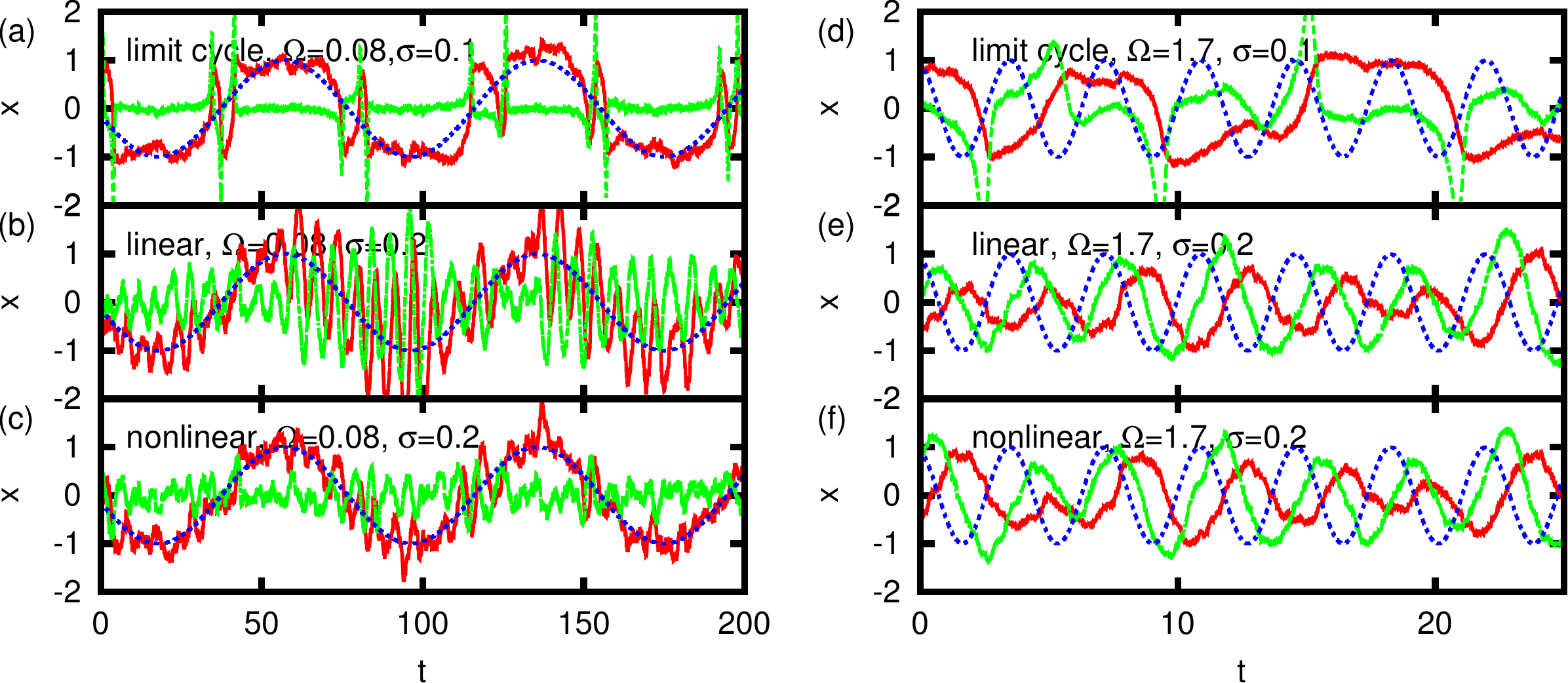}
\caption{(color online)
The time evolution of $x_1$ (solid line) and $x_2$ (dashed line)
when there is additive external forcing (dotted line, $A=1$).
The external forcing has angular frequency $\Omega=0.08$
for (a-c), and  $\Omega=1.7$  for (d-f).
(a) and (d) limit cycle oscillator with $d=2$ and $B=10$ with noise
($\sigma=0.1$).
(b) and (e) linear noise-induced oscillator with $d=-0.1$ and $B=0$ with noise
($\sigma=0.2$).
(c) and (f) nonlinear noise-induced oscillator with $d=-0.1$ and $B=1$ with noise
($\sigma=0.2$).
For the limit cycle oscillator (a and d),
the noise makes the orbit irregular,
and the phase sometime slips.
In the linear noise-induced oscillator for small external
angular frequency, we can clearly see that  the noise put the
oscillation with angular frequency close to $\omega_\ell$ on top of
one-to-one entrainment behavior (b).
When $\Omega$ is larger than $\omega_\ell$ (e), the external angular
frequency { 
is more visible, due to the smaller noise compared to the amplitude}.
The nonlinear noise-induced oscillator behaves again
very similar to the linear case in entrainment behavior
(c and f), except for the suppression of large amplitude.
}
\label{AforceExampleNoise}
\end{figure*}
The addition of noise blurs the entrainment
behavior, as depicted in Fig. \ref{AforceExampleNoise}.
For the limit cycle oscillator (Fig. \ref{AforceExampleNoise}a and d),
we can see that the noise makes the orbit irregular,
which can make the phase to slip.
In the linear noise-induced oscillator for small external
angular frequency, we can clearly see that the noise induces the
oscillation with angular frequency close to $\omega_\ell$ on top of
one-to-one entrainment behavior (Fig. \ref{AforceExampleNoise}b),
{ as expected from the auto-correlation eq. (\ref{AC})}.
When $\Omega$ is larger than $\omega_\ell$,
the external angular
frequency is { more visible, because the noise $\sigma$
is small compared to the amplitude $A$ for this case,  though both frequencies should be present}.
The nonlinear noise-induced oscillator behaves again
very similar to the linear case in entrainment behavior
(Fig. \ref{AforceExampleNoise}c and f).  
The visible difference is a suppression of large amplitude by
the nonlinear term.

\paragraph{"Devil's staircase" and "Arnold's tongues".}
\begin{figure*}[H]
\includegraphics[width=\textwidth]{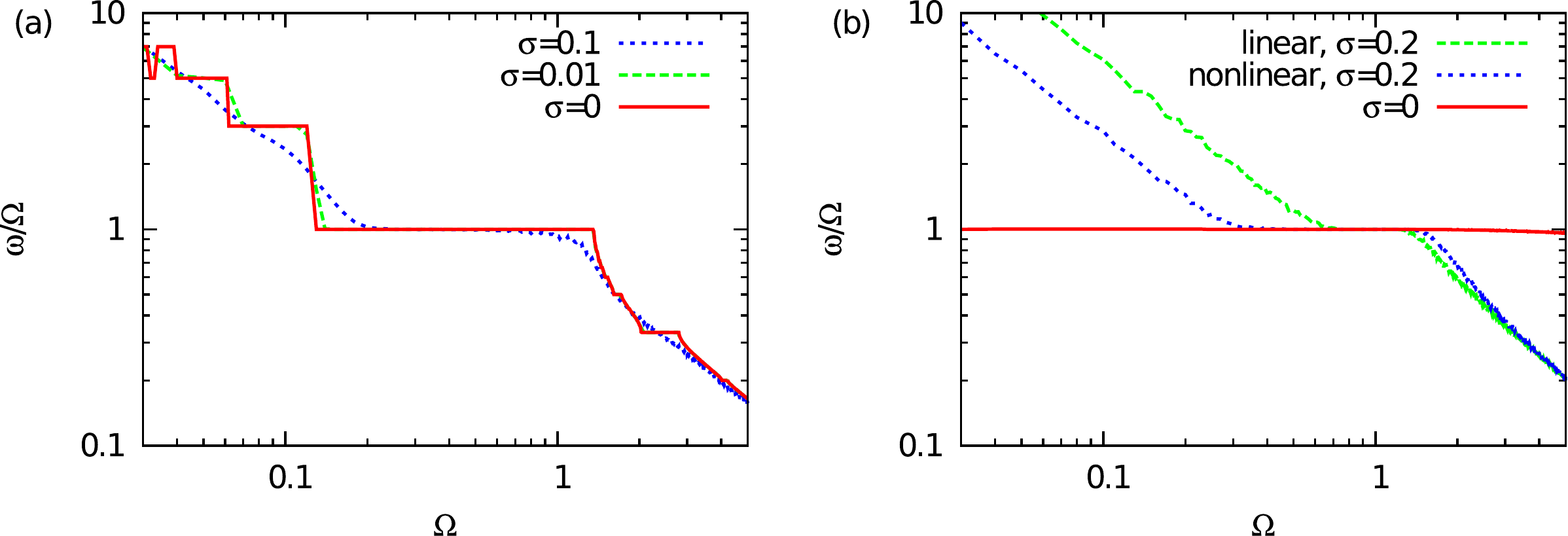}
\caption{"Devil's staircase"
for limit cycle oscillator (a) and
linear and nonlinear systems with a stable fixed point (b) under additive forcing with $A=1$.
(a) The limit cycle oscillator with $d=2$ and $B=10$
with $\sigma=0.01$ (dotted line), $\sigma=0.1$ (dashed line)
and $\sigma=0$ (solid line).
(b) The systems with a stable fixed point ($d=-0.1$).
For the case without noise $\sigma=0$ (solid line),
both linear ($B=0$) and nonlinear  ($B=1$) systems show
only one-to-one entrainment.
With noise, the noisy oscillations around the one-to-one entrained
orbit is induced, as shown with
$\sigma=0.2$ (linear case with $B=0$ is  shown by dashed line,
and nonlinear case with $B=1$ is shown by dotted line).
}\label{Figure-DSA}
\end{figure*}

For deterministic limit cycles, the
plot of $\omega/\Omega$ vs $\Omega$ for  
a fixed amplitude of external forcing shows a infinitely complex structure
with fractal nature, known as Devil's staircase \cite{JBB1, JBB2,Parlitz}.
For the present system of limit cycle oscillator without noise, this is also observed
as shown in Fig. \ref{Figure-DSA}(a) (solid line).
As noise increases, the phase slips occasionally,
therefore narrow entrainment regions become harder to recognize
(Fig. \ref{Figure-DSA}a, dashed and dotted  line).
For the systems with a stable fixed point, there is only one-to-one entrainment
for the no noise case (Fig. \ref{Figure-DSA}b,  solid line),
while noise induced oscillation around the entrained solution
will add some phase slips giving a change in the angular frequency
when the entrainment is not so strong,
resulting in an escape from the one-to-one ratio as shown in
Fig. \ref{Figure-DSA}(b).

\begin{figure*}[H]
\includegraphics[width=\textwidth]{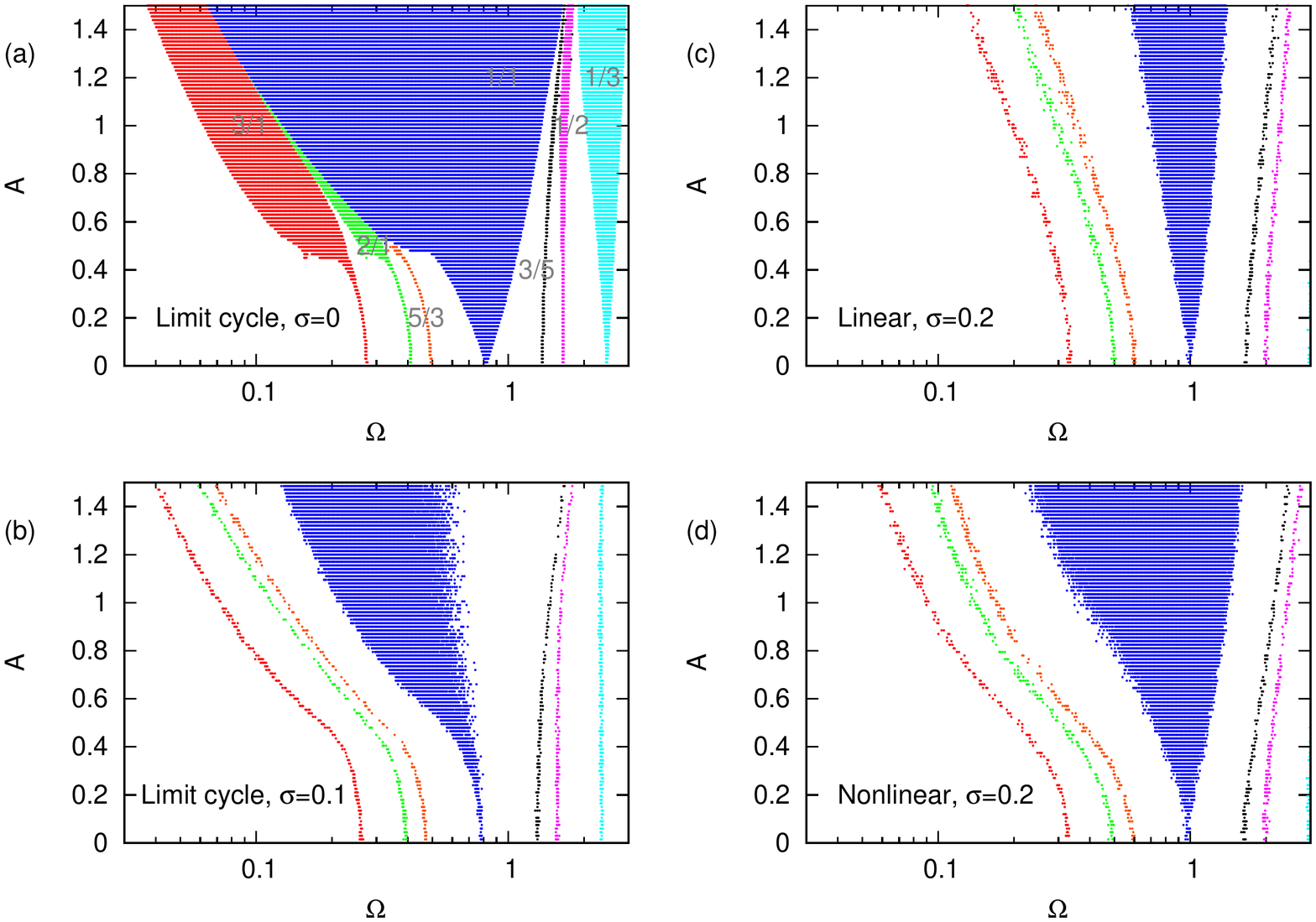}
\caption{"Arnold's tongue" with additive forcing
for limit cycle oscillator without (a) and
with (b) noise
and for noise-induced oscillator with noise
for linear (c) and nonlinear (d) case.
The horizontal axis is the external frequency
$\Omega$, and the vertical axis is the forcing amplitude $A$'
Entrainment is defined as $\omega/\Omega$ is within
1\% of the given value.
(a) The limit cycle oscillator with $d=2$ and $B=10$
with $\sigma=0.0$, which shows standard 'Arnold's tongue'.
Noise ($\sigma=0.1$) make phases to slip,
resulting in smaller region of entrainment (b).
For noise induced oscillator with noise (c: $d=-0.1, B=0, \sigma=0.2$,
d: $d=-0.1, B=1, \sigma=0.2$), the tongue-like triangle
structure is observed only for 1/1 entrainment.
}\label{Figure-TongueA}
\end{figure*}

When entrainment regions for various
values of $\omega/\Omega$ are plotted in
the $A$-$\Omega$ plain, it gives
an ``Arnold's tongue'' structure
for the deterministic limit cycles:
The entrainment regions widen as
the external forcing amplitude $A$ grows,
resulting in tongue-like shapes of the entrainment region --
when $A$ is large enough the tongues start to overlap \cite{JBB1,JBB2}.
This can be seen in the limit cycle
oscillator without noise in Fig. \ref{Figure-TongueA} (a).
When noise is added, the phase of the oscillator sometimes slips,
resulting in narrower tongues (Fig. \ref{Figure-TongueA} b).
For the noise-induced oscillators (i.e. with a stable fixed point),
there exists only 1/1 entrainment without noise,
and with noise 1/1 entrainment is the only case
that gives the tongue-like structure,
both for the linear and nonlinear cases
(Fig. \ref{Figure-TongueA} cd).
We see other ratios of entrainment ''regions'',
because for a given $A$ with changing $\Omega$,
$\omega/\Omega$ changes continuously outside of the
entrainment region (e.g. Fig.\ref{Figure-DSA} b).

\subsection{Multiplicative forcing}
\subsubsection{Linear case without noise}
We next consider the multiplicative forcing
\begin{equation}
\dot{\bi x}(t)=\mathbf L{\bi x}(t)+\mathbf M(t){\bi x}(t),
\label{multieq}
\end{equation}
where the matrix $\mathbf M(t)$ satisfies
\begin{equation}
\mathbf M(t+T)=\mathbf M(t)
\end{equation}
with $T=2\pi/\Omega$.
It is known from Floquet theory \cite{Haken} that the solution matrix
of this equation is expressed as
\begin{equation}
\mathbf{Q}(t)=e^{\mathbf{\Lambda} t}\mathbf{U}(t),
\end{equation}
where
\begin{equation}
\mathbf U(t+T)=\mathbf U(t),
\end{equation}
and a general solution is the linear combinations of
column vectors consisting of $\mathbf Q(t)$.
The eigenvalues of the matrix $\mathbf \Lambda$,
called Floquet exponents, determine the stability of the solution:
The solution will converge to the fixed point
when the real parts of the Floquet exponents
are all negative, and diverges if some Floquet exponent have
positive real parts.
Therefore, no entrainment behavior will be observed
for a linear noise-induced oscillator without noise
under multiplicative forcing.

\begin{figure}[H]
\begin{center}
\includegraphics[width=0.45\textwidth]{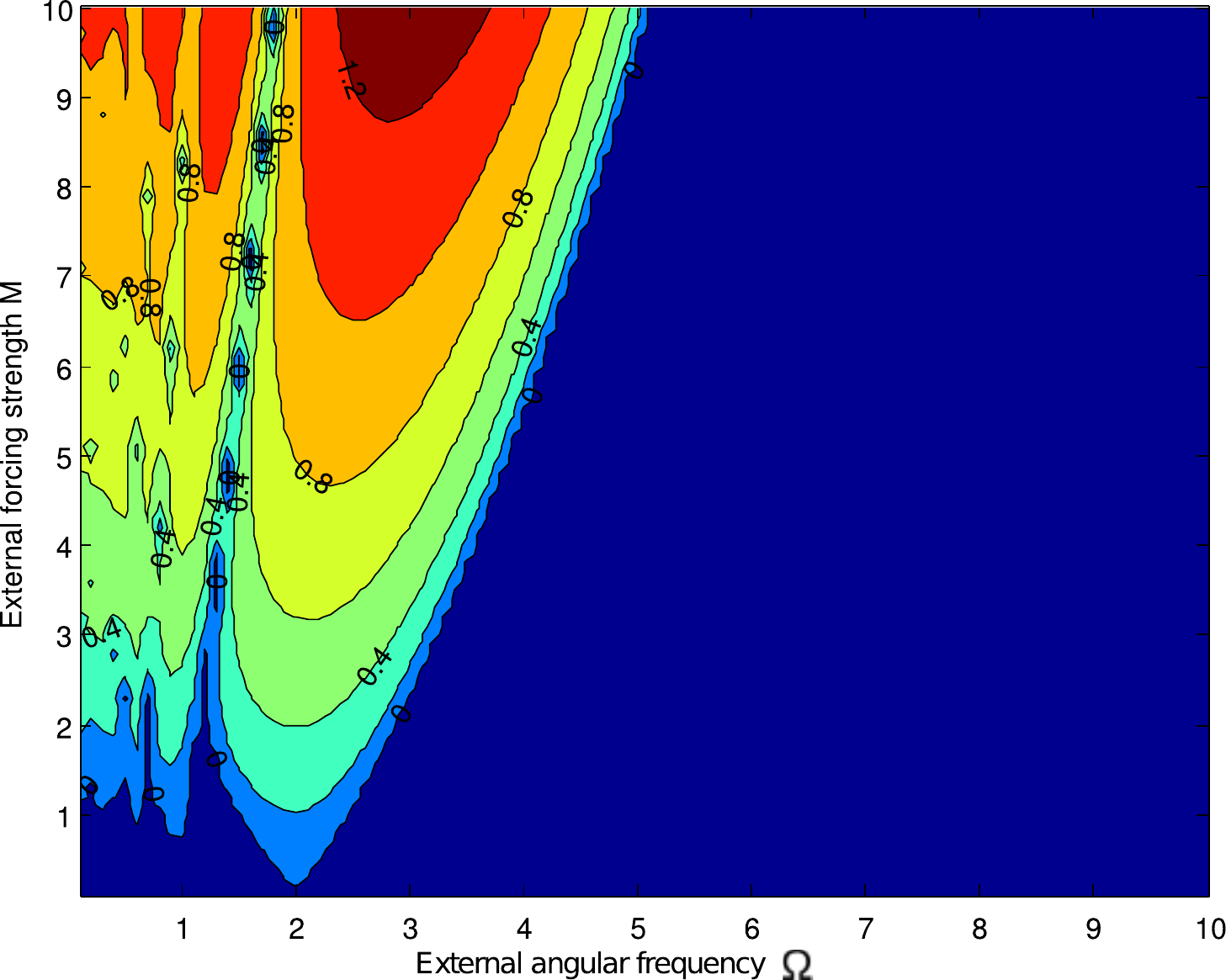}
\end{center}
\caption{Maximum real part of the
Floquet exponent
$\lambda_R$ for various $M$ and $\Omega$,
for the linear system with stable fixed point
($d=-0.1$ and $B=0$) without noise ($\sigma=0$).}
\label{Mplot}
\end{figure}

In Fig.\ref{Mplot},  we show numerically calculated
the maximum real part of the {Floquet exponents} $\lambda_R$
for (\ref{multieq}) with (\ref{whatisM})
with $d=-0.1$ and $B=0$, as a function of
amplitude of forcing $M$ and external frequency $\Omega$.
When $\lambda_R<0$ (dark blue region), $|\bi x|$
will exponentially decays to zero, otherwise
$|\bi x|$ will diverge { except for
the marginal case $\lambda_R=0$}.

\subsubsection{Numerical results}
\paragraph{Without noise.}
\begin{figure*}[H]
\includegraphics[width=\textwidth]{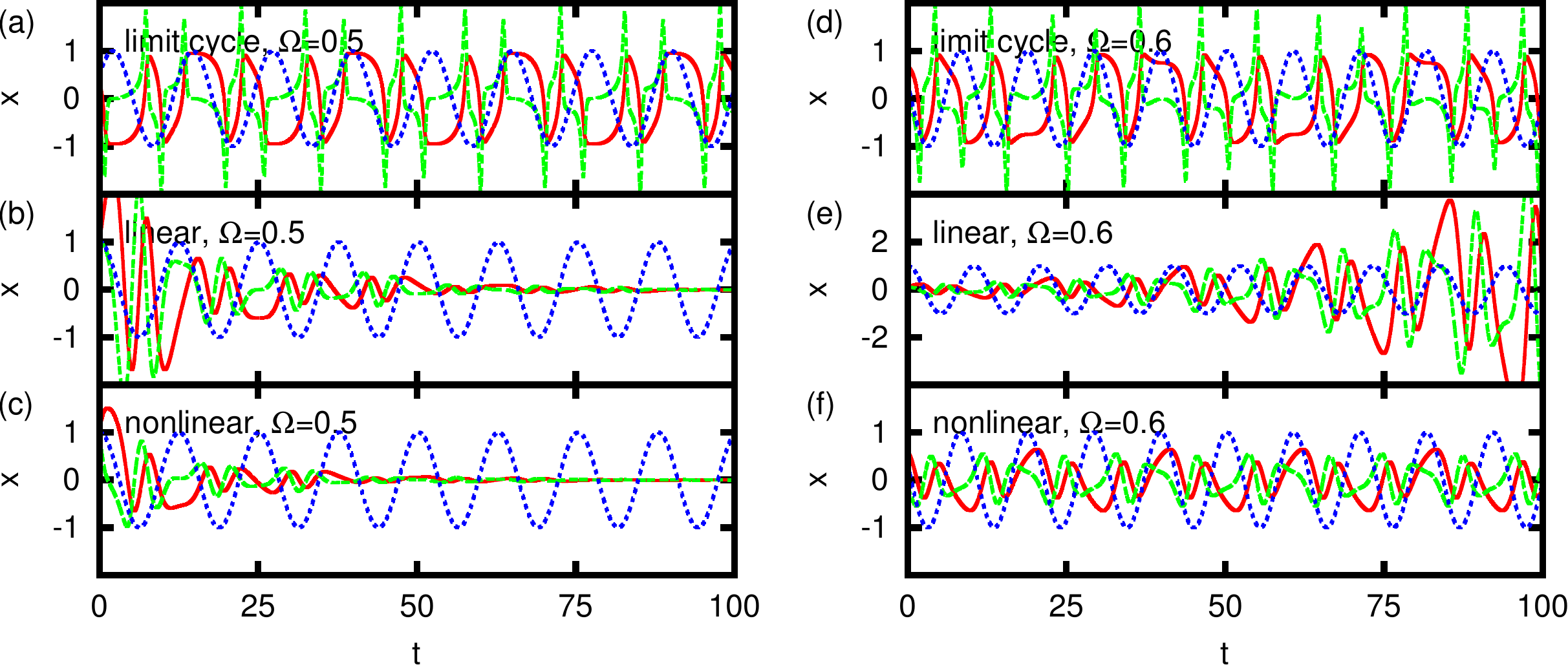}
\caption{(color online)
Without noise: The time evolution of $x_1$ (solid line) and $x_2$ (dashed line)
when there is multiplicative external forcing (dotted line, $M=1$).
The external forcing has angular frequency $\Omega=0.5$
for (a-c), and  $\Omega=0.6$  for (d-f).
(a) and (d) limit cycle oscillator with $d=2$ and $B=10$ without noise
($\sigma=0$).
(b) and (e) linear system with a stable fixed point
with $d=-0.1$ and $B=0$ without noise
($\sigma=0$). The transient behavior is shown.
Note that the y-range in (e) is different from other plots.
(c) and (f) nonlinear system
with a stable fixed point with $d=-0.1$ and $B=1$ without noise ($\sigma=0$).
The limit cycle oscillator shows entrainments (a,d), but the linear
system either decays to zero (b) or diverges (e).
The nonlinear system either decays (c) or entrains (f).
}
\label{MforceExample}
\end{figure*}

Figure \ref{MforceExample} shows the entrainment behaviors for
multiplicative forcing.
For the limit cycle oscillator (Fig. \ref{MforceExample} ac),
there is no qualitative difference from the additive noise case,
i.e., the system shows entrainment with various frequency
ratio $\omega/\Omega=P/Q$.
For the linear system with a stable fixed point without noise,
on the other hand, the system can either
decay to the fixed point (Fig. \ref{MforceExample} b)
or diverge (Fig. \ref{MforceExample} d),
which can be predicted from the Flouquet exponents (Fig.\ref{Mplot}).
When nonlinear term is added,
it does not prevent the decay
(Fig. \ref{MforceExample} c),
but the divergent behavior is suppressed
and system shows the entrainment behavior (Fig. \ref{MforceExample} f).
The frequency ratio $\omega/\Omega$
is not necessarily 1/1;
the example in Fig. \ref{MforceExample}(f) gives
$\omega/\Omega=3/2$.
\paragraph{With noise.}
\begin{figure*}[H]
\includegraphics[width=\textwidth]{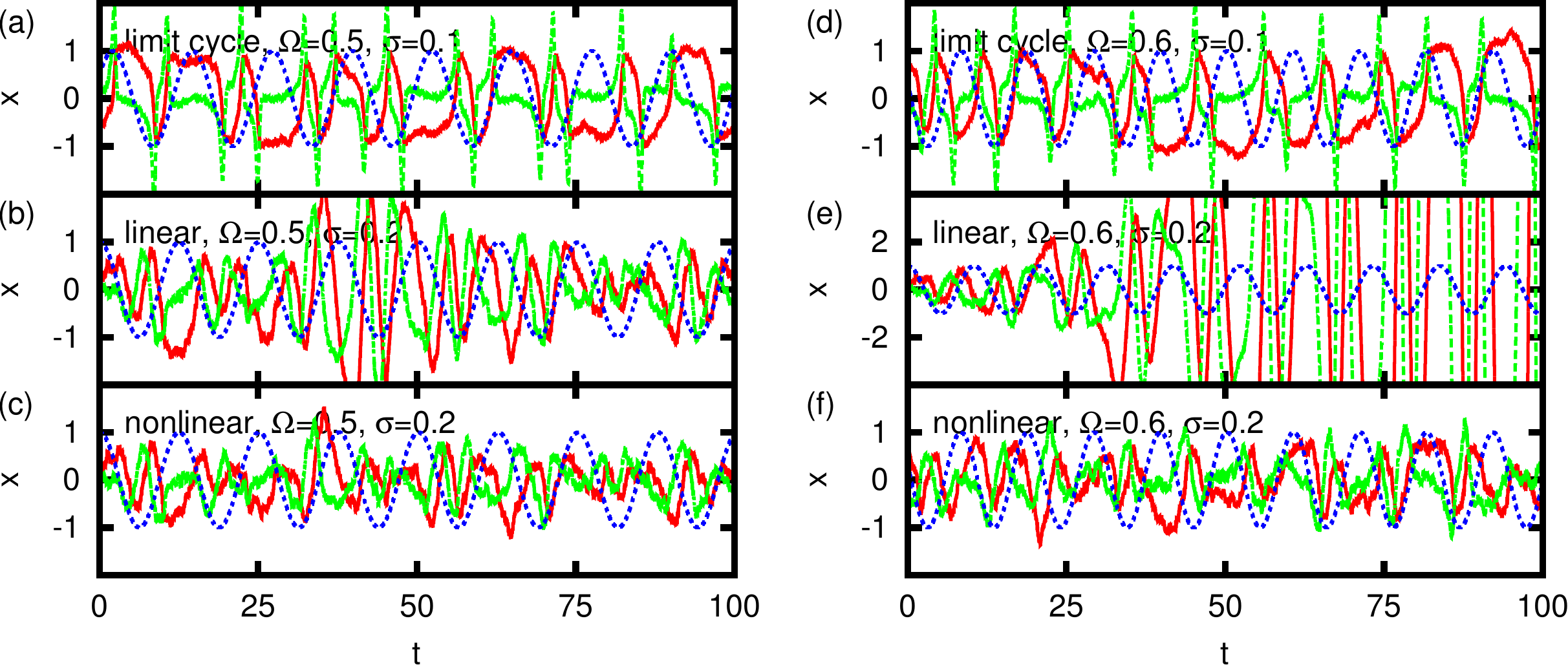}
\caption{(color online)
With noise: The time evolution of $x_1$ (solid line) and $x_2$ (dashed line)
when there is multiplicative external forcing (dotted line, $M=1$).
The external forcing has angular frequency $\Omega=0.5$
for (a-c), and  $\Omega=0.6$  for (d-f).
(a) and (d) limit cycle oscillator with $d=2$ and $B=10$ with noise
($\sigma=0.1$).
(b) and (e) linear noise-induced
oscillator with $d=-0.1$ and $B=0$ with noise
($\sigma=0.2$). Note that the y-range in (e) is different from other plots.
(c) and (f) nonlinear noise-induced oscillator with $d=-0.1$ and $B=1$ with noise ($\sigma=0.2$).
The limit cycle oscillator shows entrainments with some phase slips
(a,d). For the linear and nonlinear system,
the noise induces the oscillatory behavior,
for the parameters where the system would decay without noise (b,c).
}
\label{MforceExampleNoise}
\end{figure*}

When noise is added, the behavior changes drastically
in the noise-induced oscillators, as shown in.
Fig. \ref{MforceExampleNoise}.
The noise can induce the oscillation with
the angular frequency close to $\omega_\ell$ for the
case where the no-noise system would decay to the fixed point
(Fig. \ref{MforceExampleNoise} b, c).
On the other hand, in the linear noise-induced oscillator,
adding noise does not prevent divergence
(Fig. \ref{MforceExampleNoise} e).
For the parameters where
no-noise system would entrain, the
noise blurs the entrainment due to
occasional phase slip
for both limit cycle oscillator (Fig. \ref{MforceExampleNoise} ac)
and nonlinear noise-induced oscillator
(Fig. \ref{MforceExampleNoise} f).

\paragraph{"Devil's staircase" and "Arnold's tongue".}
\begin{figure*}[H]
\includegraphics[width=\textwidth]{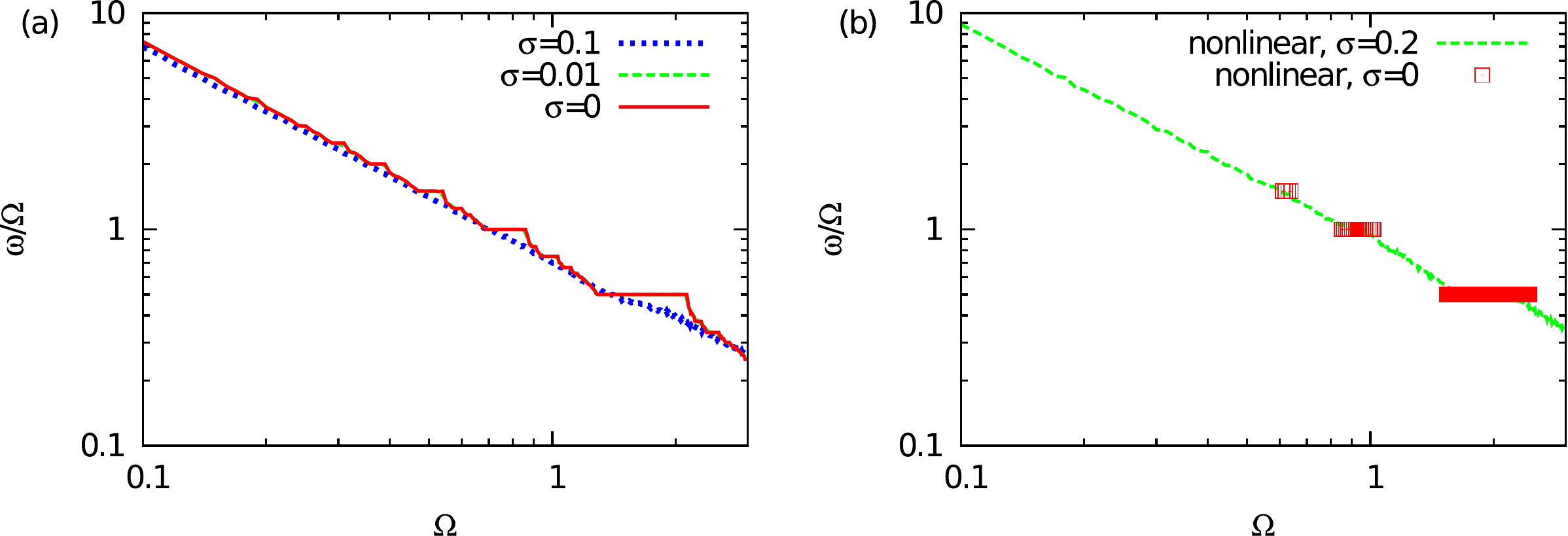}
\caption{"Devil's staircase"
for limit cycle oscillator (a) and
nonlinear noise-induced oscillator (b) under multiplicative forcing with $M=1$.
(a) The limit cycle oscillator with $d=2$ and $B=10$
with $\sigma=0.01$ (dotted line), $\sigma=0,1$ (dashed line)
and $\sigma=0$ (solid line).
(b) The nonlinear system with a stable fixed point with $d=-0.1$
and $B=1$.
For the case without noise $\sigma=0$ (solid line),
the decaying region where $\bi x$ goes to
the fixed point is not shown,
resulting in three discrete entrainment region.
With noise, oscillation is induced in the decaying regime also,
resulting in continuous line as shown for $\sigma=0.2$ (dashed line).
}\label{Figure-DSM}
\end{figure*}
\begin{figure*}[H]
\includegraphics[width=\textwidth]{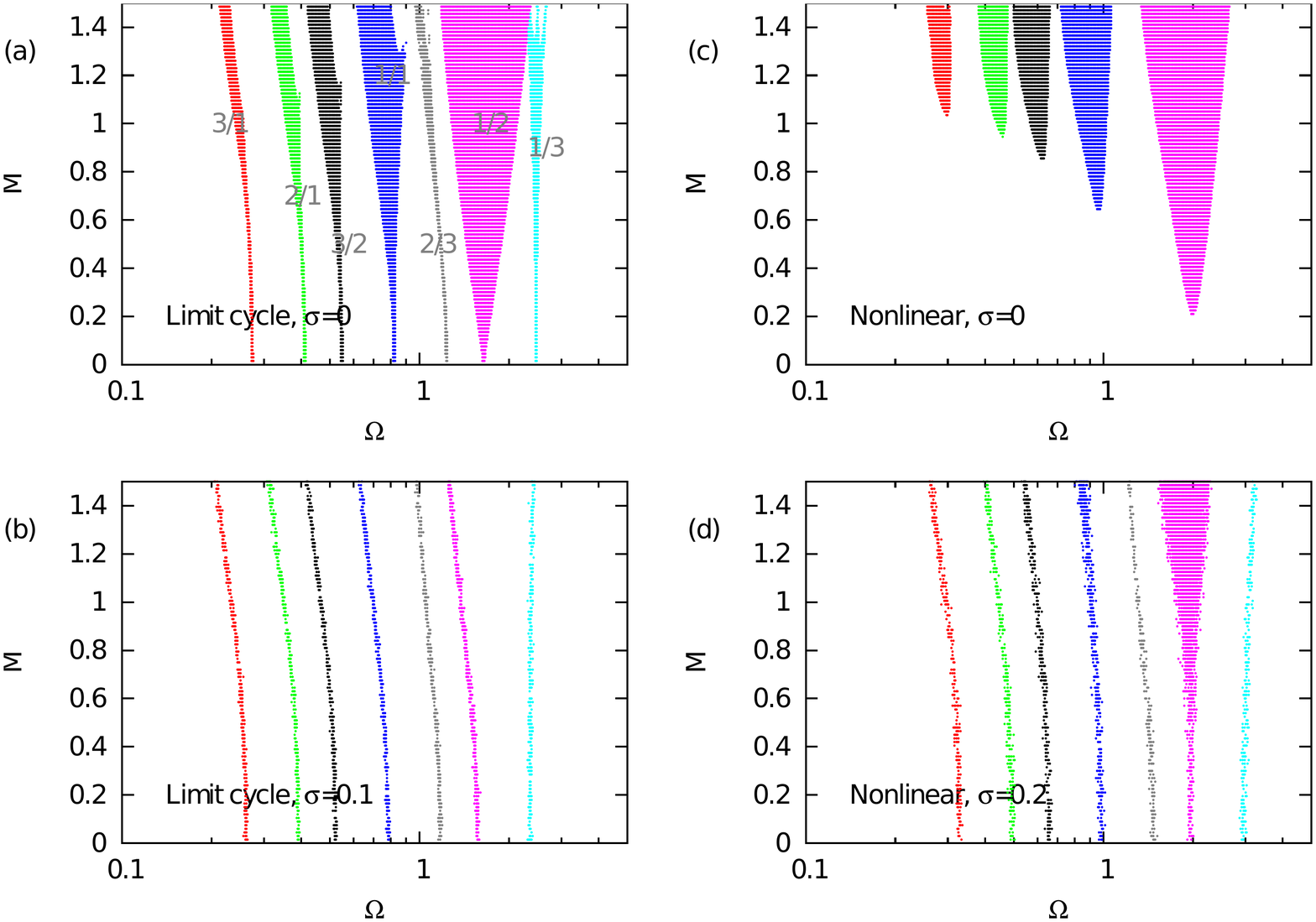}
\caption{"Arnold's tongue" with multiplicative forcing
for limit cycle oscillator without (a) and
with (b) noise
and for nonlinear system with a stable fixed point without
(c) and with noise (d).
The horizontal axis is the external frequency
$\Omega$, and the vertical axis is the forcing amplitude $M$.
Entrainment is defined as $\omega/\Omega$ is within
1\% of the given value.
(a) The limit cycle oscillator with $d=2$ and $B=10$
with $\sigma=0$ shows standard 'Arnold's tongue', while
the noise ($\sigma=0.1$) makes the region of entrainment smaller (b).
For nonlinear noise induced oscillator ($d=-0.1, B=1$; c),
there are a few entrainment regions for
no noise case ($\sigma=0$), but not all the ratios are observed.
For (c), the exponentially decaying case were excluded numerically by
the following way: The equations are integrated with initial condition
$x(1)=1$ and $x(2)=0$, and if the average amplitude
for $390 \pi/ \Omega<t<400 \pi/ \Omega$  is less than 90\% of
the average amplitude for $200 \pi/ \Omega<t<210 \pi/ \Omega$,
then the solution is excluded.
}\label{Figure-TongueM}
\end{figure*}
We also study the "Devil's staircase"
for the multiplicative forcing.
For the limit cycle oscillator without noise, we again see proper devil's
staircase, where noise will blur the entrainment behaviors
(Fig.\ref{Figure-DSM} a).
For the noise-induced oscillators,
only the nonlinear case is studied because
the linear case may diverge depending on the parameter values.
Without noise, we see discrete finite regions of entrainment
(Fig.\ref{Figure-DSM} b squares),
while noise induces the oscillations in the decaying region resulting in a
continuous line
(Fig.\ref{Figure-DSM} b dashed line).

The Arnold's tongue structure for the
limit cycle is similar to those in the
additive forcing case, as seen in Fig.\ref{Figure-TongueM}(a) and (b).
The Arnold's tongues for all the entrainment ratios are
observed without noise, and noise makes the regions smaller.
For the nonlinear system with a stable fixed point without noise,
there are entrainment regions for a few rational
ratios, but the ones that appear are problem specific -
for instance, in the present case, the $\omega/\Omega=1/3$ is not observed at
all in Fig.\ref{Figure-TongueM}(c).
With noise (Fig.\ref{Figure-TongueM}d),
the entrainment regions shrinks,
but at the same time  the system can occasionally pass
the given ratio of $\omega/\Omega$, resulting in
narrow line of ``fake'' entrainment.

\section{Biological Example:
Entrainment of TNF-driven NF-$\kappa$B system}
{
In this section, we study a biological example,
TNF-driven NF-$\kappa$B system,
to confirm the generality of the
entrainment behavior for the limit cycle system under weak noise.
}

{
The system has been studied for the deterministic case in ref.\cite{JK}.
NF-$\kappa$B is a transcription factor, and it has been 
verified experimentally that
NF-$\kappa$B level in the nucleus shows sharp oscillations
after treatment with tumor-necrosis factor (TNF)\cite{TNF1,TNF2}.
The interaction network involves a negative feedback loop
between the NF-$\kappa$B and an inhibitor, IkB$\alpha$,
which is the main mechanism for the oscillations. TNF
modulates the state of the IkB$\alpha$ and hence affects the oscillation.
TNF can be added externally to the cell, therefore it can serve as
a possible probe to study the entrainment, i.e., we can use
TNF level as the external forcing term.
In ref.\cite{JK}, the system was modeled by 5 dimensional coupled nonlinear
ODEs,
\[
\frac{d \bi x}{dt}=\bi F(\bi x, [TNF]),
\]
where
\begin{eqnarray}
F_1&=&
k_{Nin}(N_{tot}-x_1)\frac{K_I}{K_I+x_3}
-k_{lin}x_3\frac{x_1}{K_N+x_1}, \\
F_2&=&
k_t x_1^2-\gamma_m x_2,\\
F_3&=&
k_{tl}x_2-\alpha x_4(N_{tot}-x_1)\frac{x_3}{K_I+x_3},\\
F_4&=&
k_a [TNF](I_{tot}-x_4-x_5)-k_ix_4,\\
F_5&=&k_ix_4
-k_px_5\frac{k_{A20}}{k_{A20}+A_{20}[TNF]}.\label{x5}
\end{eqnarray}
The variable $x_1$ denotes the nuclear NF-$\kappa$B level,
and $[TNF]$ denotes the TNF level, which we shall change to a periodically
external forcing onto the system.
The biological meaning of the
variables and parameter values are summarized in Table \ref{parameters}.
Note that $[TNF]$ appears twice in eq. (\ref{x5})
in the terms multiplied with $\bi x$, therefore this is an example of
multiplicative forcing.
We study this system with adding a Gaussian white noise in each term,
i.e.,
\[
\frac{d \bi x}{dt}=\bi F(\bi x, [TNF])+\sigma \bi \Gamma.
\]
}
\begin{table}[H]
\caption{\label{parameters} Variables and the parameters in the
TNF-driven NF-kB oscillation, from ref.\cite{JK}.
}\footnotesize\rm
\begin{tabular}{ll}
\hline
$x_1$ & nuclear NF-kB level\\
$x_2$ & IkB mRNA level\\
$x_3$ & cytoplasmic IkB protein level\\
$x_4$ & active IKK level\\
$x_5$ & inactive IKK level\\
\hline
$N_{tot}$ & Total NfkB level, 1 $\mu$M\\
$I_{tot}$ & Total IKK level, 2.0 $\mu$M\\
\hline
$k_{Nin}$ & 5.4 min$^{-1}$\\
$K_I$ & 0.035 $\mu$M\\
$k_{lin}$ & 0.018 min$^{-1}$\\
$K_N $& 0.029 $\mu$M\\
$k_t$ & 1.03 $\mu$M$^{-1}$ min$^{-1}$\\
$\gamma_m$ & 0.017 min$^{-1}$\\
$k_{tl}$ & 0.24 min$^{-1}$\\
$\alpha$ &1.05 $\mu$M$^{-1}$\\
$k_a$ & 0.24 min$^{-1}$\\
$k_i$ & 0.18 min$^{-1}$\\
$k_p$ & 0.036 min$^{-1}$\\
$k_{A20}$ & 0.0018 $\mu$M\\
$A_{20}$ & 0.0028 $\mu$ M\\
\hline
\end{tabular}
\end{table}

\begin{figure}[H]
\includegraphics[width=\textwidth]{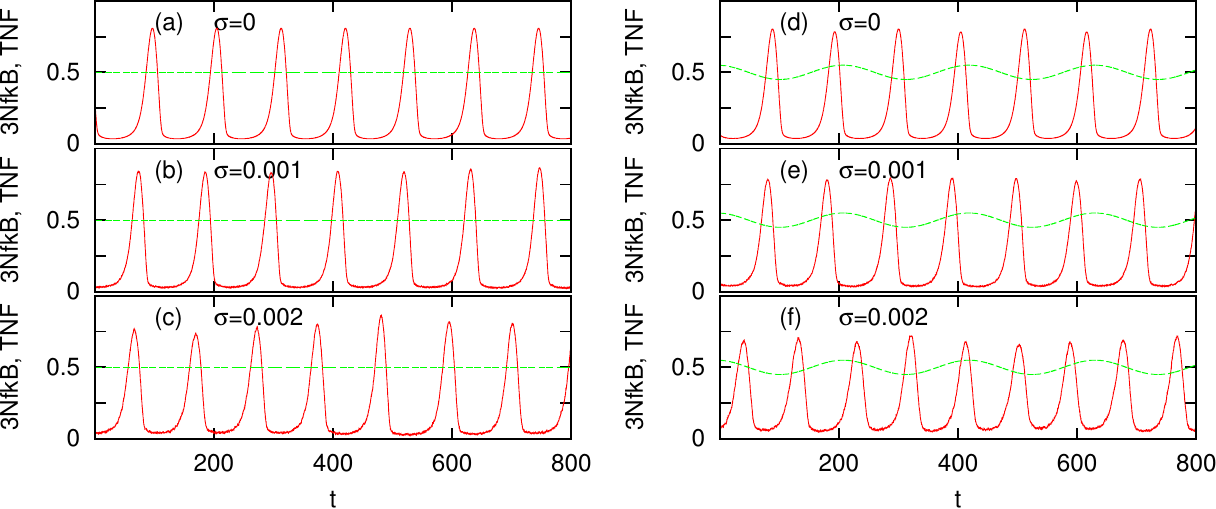}
\caption{The oscillations and entrainment for TNF-driven
NF-$\kappa$B system. Spontaneous oscillations
with [TNF]=0.5 for (a) no noise ($\sigma=0$) case
and (b)$\sigma=0.001$, (c)$\sigma=0.002$.
}\label{NFkBexamples}
\end{figure}

{
Figure~\ref{NFkBexamples}(a-c) shows the spontaneous oscillation
of nuclear NF-$\kappa$B, when $[TNF]$ is kept constant at $[TNF]=0.5$,
without (a) and with noise (b-c). We see clear periodic oscillation with
the period around $110$ (minutes).
We then modulate the $[TNF]$ level around this basal level \cite{JK} as
\begin{equation}
[TNF]=0.5+M_{TNF}\sin(\Omega t).
\end{equation}
This has been studied in the no-noise case
by Jensen and Krishna~\cite{JK}, 
and it was found that the entrainments of various
ratios can occur, when the frequency of the NF-$\kappa$B level
is determined based on the frequency of the peaks.
Figure ~\ref{NFkBexamples}(c) shows an example of 1/2 entrainment,
for $M_{TNF}=0.05$ and $\Omega=0.0297$, in the deterministic case.
With weak enough noise, the entrainment is
maintained (Fig.~\ref{NFkBexamples}e),
but larger noise induces phase slips (Fig.~\ref{NFkBexamples}f),
as has been seen in the Van der Pol system.
}

\begin{figure}[H]
\includegraphics[width=\textwidth]{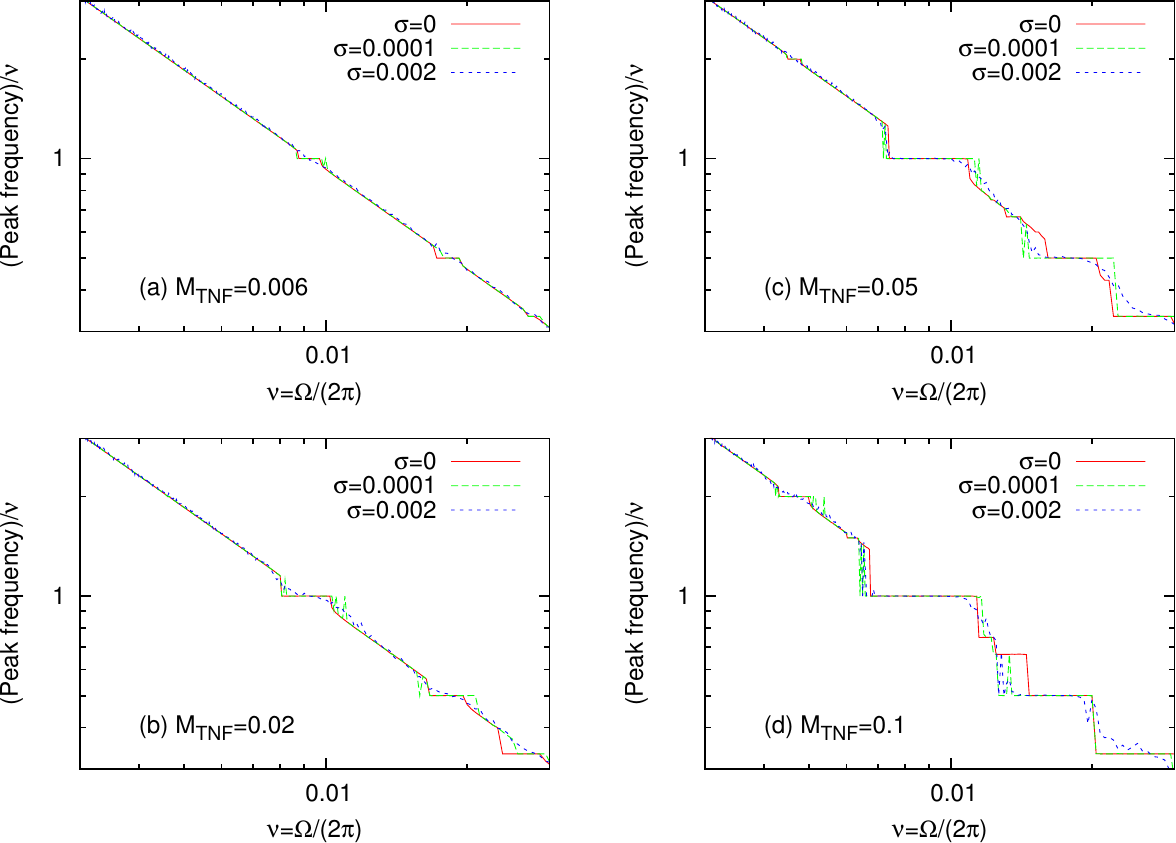}
\caption{``Devil's staircase'' for TNF-driven
NF-$\kappa$B system without and with noise, with (a)$M_{TNF}=0.006$,
(b)$M_{TNF}=0.02$, (c)$M_{TNF}=0.05$, and
(d)$M_{TNF}=0.1$.
The entrainment regions are calculated from the frequency of the peaks in the
the deterministic case. In the finite noise case, we
define the nuclear NF-$\kappa$B peak as follows:
We first determined the maximum value $N_{max}$ and
the minimum value $N_{min}$ of
$x_1$ of the steady state in the deterministic simulation
for the given parameters.
We then calculate two thresholds, $N_{H}=(N_{max}+N_{min})/2$
and $N_{L}=(N_{max}+3N_{min})/4$.
Next we perform the corresponding simulation with finite $\sigma$.
We define a switching event from the ``low'' state to
``high'' state when $x_1$ exceeds $N_H$, while
the reverse switching happens when 
$x_1$ becomes smaller than $N_L$.
The number of peaks are calculated from how often the ``high'' states
are reached.
This way we can filter out the wiggly motion
due to the noise and thus define the overall peak.
}
\label{NfkBstairs}
\end{figure}

{
In Fig.~\ref{NfkBstairs}, several Devil's staircases are
shown with and without noise. In Ref.~\cite{JK} 
the Arnold tongues have been calculated, and it has been
demonstrated that general P/Q entrainments occur.
A characteristic observation to this system is that the tongues overlap
easier for larger external frequency;
e.g., $M_{TNF} \approx 0.04$ for 1/3 and 1/2 tongues
to overlap, while the 2/1 and 5/2 tongues do not overlap
even at $M_{TNF}=0.1$.
When the Devil's staircase are calculated for overlapping region,
non-smooth or irregular jumps between the steps can be seen,
and is thus dependent on initial conditions in general.
This is visible in our data in Fig.\ref{NfkBstairs},
for large $M_{TNF}$ and larger $\Omega$.
When weak noise is added, it enables the system to
jump to other overlapping tongues, which results in
irregular behavior around the entrainment regions.
As the noise becomes larger, the entrainment is again smoothed away
by phase slips.
}

\section{Summary and Discussion}
\begin{table}[H]
\caption{\label{tablesummary} Summary of entrainment behavior of oscillators
under additive and multiplicative forcing.
A and M in the "force" column represent additive and multiplicative forcing,
respectively.
}\footnotesize\rm
\begin{tabular*}{\textwidth}{@{}l*{15}{@{\extracolsep{0pt plus12pt}}l}}
\hline
        Oscillator & No noise & With noise & Force \\ \hline \hline
Limit cycle & entrainment to any $P/Q$ &
entrainment to any $P/Q$ with phase slips
& A \\ \cline{2-4}
& entrainment to any $P/Q$ &
entrainment to any $P/Q$ with phase slips
& M \\ \hline
Linear  & one-to-one entrainment$^*$
& one-to-one entrainment$^*$ with phase slips
& A \\ \cline{2-4}
noise-induced  & decay or diverge & noise-induced oscillation with
$\sim \omega_\ell$
 & M \\
 && or diverge &\\
 \hline
Nonlinear  & one-to-one entrainment$^*$ & one-to-one entrainment$^*$ with phase slips
& A \\ \cline{2-4}
noise-induced & small decay or &
noise-induced oscillation with $\sim \omega_\ell$ or  & M  \\
&some $P/Q$ entrainment&
some $P/Q$ entrainment with phase slips  &  \\ \hline
\end{tabular*}
\begin{itemize}
\item[]  {$^*$ All the frequencies contained in the forcing can be observed. }
\end{itemize}
\end{table}\normalsize

Our motivation behind this work was to ask: Can one by
applying an external periodic forcing and studying entrainment determine whether
an oscillating system is driven by a linear mechanism (noise induced oscillator)
or a non-linear mechanism (limit cycle oscillator)?
Our answer to this question is {
generally yes}.
Our obtained results on entrainment behavior of oscillators
are summarized in Table \ref{tablesummary}.

When the forcing is additive, there is clear difference
between the limit cycle oscillators and
the noise-induced oscillators.
The former can entrain to any frequency ratio,
while the latter shows only one-to-one entrainment.
Therefore, if one see entrainment to $P/Q\ne 1$ ratio,
under additive forcing, it is a
sign of limit cycle oscillator.

When the forcing is multiplicative,
the non-linear noise-induced oscillators can
also show $P/Q\ne 1$ ratio entrainment, { but not necessarily all of the rational ratios}.
If the system is noise-induced oscillator and
the non-linear term is small,
one might be able to capture the diverging tendency
of the amplitude, because saturation happens when
the amplitude is large enough to make
the non-linear term relevant.
In such a case, one might see big difference in
amplitude for a fixed $M$ with varying $\Omega$.

We thus urge experiment to be performed on oscillating biological systems.
It is well known that some proteins (p53, NF-kB, Wnt) can oscillate in cells
under stress responses. In the case of p53, both non-linear \cite{Tiana,Geva-Zatorsky06}
and linear models have been proposed \cite{Geva-Zatorsky}.
By applying an external
time dependant signal such as
DNA damaging radiation or drugs which specifically perturb the p53 circuit,
it might be possible to entrain the internal
oscillation and draw conclusions on the basis of our results summarized
in Table \ref{tablesummary}. In the case of NF-kB oscillations, one
might be able to entrain the internal oscillation by an externally
varying cytokine (like TNF) signal \cite{JK}. Potentially, it could lead
to a way of controlling the DNA-repair pathway.

{
The present research also opens for further theoretical investigations.
The $P/Q\ne 1$ entrainment of the winding number for a nonlinear noise-induced oscillator
with multiplicative forcing
is a purely numerical observation, and further research is needed to
refine the condition when this can occur.
We did not study the strong noise case either, and it would be
interesting to investigate in more details the active role of noise in the entrainments of limit cycles.
In many biological examples, where dynamics are molecular reaction based, additive Gaussian white noise is not appropriate for large noise because it does not reflect the noise amplitude dependence on the molecule number: instead either a concentration dependent noise amplitude or a stochastic treatment of molecule numbers should be performed. 
}

{ Finally, we would like to briefly comment on  "noise-induced" oscillations
by mechanisms other than the linear model studied here.
It has been long known that, when noise is added to excitable system with
a stable fixed point, regular oscillatory behaviour can be observed
at a certain level of noise (coherence resonance)
\cite{Gang,Pikovsky}.
Since the nonlinearity plays an  important role in an oscillation,
such a system shows mode-locking behaviour similar to
the deterministic nonlinear oscillators \cite{Zhou}.
More recently, in gene network models with negative feedback,
it has been shown that
the noise due to finiteness of the number of molecules
can modify the condition for oscillatory behaviour \cite{Loinger}
or enhance the oscillation \cite{Morant}.
It would also be interesting to see the entrainment behaviour
in such systems.
}
\section*{Acknowledgement}
This study was supported by the Danish National Research Foundation through the Center for Models of Life.

%\section*{References}

\end{document}